\documentclass[aps,pre,twocolumn,superscriptaddress,showpacs]{revtex4}
\usepackage{amsfonts}
\usepackage{amsmath}
\usepackage{amssymb}
\usepackage{graphicx}
\usepackage{epsfig}

\input{tcilatex}

\begin{document}

\title{Spectral fluctuations of billiards with mixed dynamics: from time
series to superstatistics}
\author{A. Y. Abul-Magd}
\affiliation{Department of Mathematics, Faculty of Science, Zagazig University, Zagazig,
Egypt}
\affiliation{Faculty of Engineering, Sinai University, El-Arish, Egypt}
\author{B. Dietz}
\affiliation{Institut f\"{u}r Kernphysik, Technische Universit\"{a}t Darmstadt, D-64289
Darmstadt, Germany}
\author{T. Friedrich}
\affiliation{Institut f\"{u}r Kernphysik, Technische Universit\"{a}t Darmstadt, D-64289
Darmstadt, Germany}
\author{A. Richter}
\affiliation{Institut f\"{u}r Kernphysik, Technische Universit\"{a}t Darmstadt, D-64289
Darmstadt, Germany}

\begin{abstract}
A statistical analysis of the eigenfrequencies of two sets of
superconducting microwave billiards, one with mushroom-like shape and the
other from the familiy of the Lima\c{c}on billiards, is presented. These
billiards have mixed regular-chaotic dynamics but different structures in
their classical phase spaces. The spectrum of each billiard is represented
as a time series where the level order plays the role of time. Two most
important findings follow from the time-series analysis. First, the spectra
can be characterized by two distinct relaxation lengths. This is a
prerequisite for the validity of the superstatistical approach which is
based on the folding of two distribution functions. Second, the shape of the
resulting probability density function of the so-called superstatistical
parameter is reasonably approximated by an inverse $\chi^2$ distribution.
This distribution is used to compute nearest-neighbor spacing distributions
and compare them with those of the resonance frequencies of billiards with
mixed dynamics within the framework of superstatistics. The obtained spacing
distribution is found to present a good description of the experimental ones
and is of the same or even better quality as a number of other spacing
distributions, including the one from Berry and Robnik. However, in contrast
to other approaches towards a theoretical description of spectral properties
of systems with mixed dynamics, superstatistics also provides a description
of properties of the eigenfunctions. Indeed, the inverse $\chi^2$ parameter
distribution is found suitable for the analysis of experimental resonance
strengths in the Lima\c{c}on billiards within the framework of
superstatistics.
\end{abstract}

\date{\today}
\maketitle

\section{Introduction}

Integrable Hamiltonian dynamics is characterized by the existence of as many
conserved quantities as degrees of freedom. Each trajectory evolves on an
invariant hyper-torus in the phase space \cite{lichtenberg,chandre}. In
contrast, chaotic systems are ergodic; almost all orbits fill the energy
shell in a uniform way. Physical systems with integrable and fully chaotic
dynamics, respectively, are, however, exceptional. A typical Hamiltonian
system shows a phase space in which regions of regular motion and chaotic
dynamics coexist. These systems are known as mixed systems. Their dynamical
behavior is by no means universal, as is the case for fully regular and
fully chaotic systems. If we perturb an integrable system, most of the
periodic orbits on tori with rational frequencies disappear. However, some
of these orbits persist. Elliptic periodic orbits appear surrounded by
islands. They correspond to librational motions around these periodic orbits
and reflect their stability. The Kolmogorov-Arnold-Moser (KAM) theorem
states that invariant tori with a sufficiently incommensurate frequency
vector are stable with respect to small perturbations. Numerical simulations
show that when the perturbation increases more and more tori are destroyed.
For large enough perturbations, there are locally no tori in the considered
region of phase space. The break-up of invariant tori leads to a loss of
stability of the system, that is, to chaos. There are three main scenaria of
transition to global chaos in finite-dimensional (nonextended) dynamical
systems, one via a cascade of period-doubling bifurcations, a Lorenz
system-like transition via Hopf and Shil'nikov bifurcations, and the
transition to chaos via intermittency \cite{eckmann,bunimovich}. It is
natural to expect that there are other (presumably many more) such scenaria
in extended (infinite-dimensional) dynamical systems.

In quantum mechanics, the specification of a wave function is always related
to a certain basis. In integrable systems the eigenbasis of the Hamiltonian
is known in principle. In this basis, each eigenfunction has just one
component. That obviously indicates the absence of complexity. In the nearly
ordered regime, mixing of quantum states belonging to adjacent levels can be
ignored and the energy levels are uncorrelated. The level-spacing
distribution is well described by that for random numbers generated by a
Poissonian process, $\exp(-s)$, where $s$ is the spacing between adjacent
energy levels rescaled to unit mean spacing $D$. For a Hamiltonian with a
chaotic classical limit, on the other hand the wavefunction components are
on average uniformly distributed over the whole basis. Berry \cite{berry}
conjectured that the wavefunctions of chaotic quantum systems can be
represented as a formal sum over elementary solutions of the Laplace
equation in which the real and imaginary parts of the coefficients are
independent identically-distributed Gaussian random variables with zero mean
and variance computed from the normalization. Bohigas et al. \cite{bohigas}
put forward a conjecture (strongly supported by accumulated numerical and
experimental evidence and also by recent advances towards a proof of this
conjecture \cite{haakestuff}) that the spectral statistics of chaotic
systems follow random-matrix theory (RMT, see \cite{mehta,guhr}). The
properties of a chaotic Hamilton operator can thus be modeled by an ensemble
of random Hermitian matrices $H$ that belongs to one of three universality
classes, either the orthogonal, the unitary or the symplectic one and is
called Gaussian orthogonal (GOE), unitary (GUE) and symplectic (GSE)
ensemble, respectively. The theory is based on two main assumptions: the
matrix elements are independent identically-distributed random variables and
their distribution is invariant under unitary transformations. This leads to
the Gaussian probability density distribution for the matrix elements 
\begin{equation}
P\left( H\right) =\frac{1}{Z(\eta)}\exp\left[ -\eta\text{Tr}\left(
H^{\dagger}H\right) \right] ,  \label{GE}
\end{equation}
where $Z(\eta)=\int\exp\left[ -\eta\text{Tr}\left( H^{\dagger}H\right) %
\right] d\eta$ is the normalization constant. The Gaussian distribution is
also obtained by maximizing the Shannon entropy with the constraints of
normalization and existence of the expectation value of Tr$\left(
H^{\dagger}H\right)$, see, e.g. \cite{mehta,balian}. Information about the
statistical properties of the eigenvalues and/or eigenvectors of the matrix $%
H$ can be obtained by integrating over the undesired variables. There is
strong evidence by now, that indeed the spectral correlation functions of a
chaotic system are well decribed by those obtained from Eq.~(\ref{GE}) and
determined solely by the global symmetries of the system such as
time-reversal invariance and the value of the spin. Among the measures
representing spectral correlations, the nearest-neighbor level-spacing
distribution (NNSD) $p(s)$ has been studied extensively so far. For the
random matrix ensembles Eq.~(\ref{GE}) it is well approximated by the
Wigner-Dyson distribution, namely $p_{\beta}(s)=a_{\beta}s^{\beta}\exp(-b_{%
\beta}s^{2})$, where $\beta$ (= 1, 2, and 4 for the orthogonal, the unitary,
and the symplectic ensembles, respectively) characterizes the universality
classes. The coefficients $a_{\beta}$\ and $b_{\beta}$\ are determined by
the normalization conditions $\int_{0}^{\infty}p_{\beta}(s)ds=\int_{0}^{%
\infty}sp_{\beta}(s)ds=1$, as $a_{1}=\pi/2,~a_{2}=32/\pi^{2},~a_{4}=%
\pi^{3}2^{18}/3^6,~b_{1}=\pi/4,~b_{2}=4/\pi,$and$~b_{4}=64/9\pi$. For $s\ll1$%
, the distribution function is proportional to $s^{\beta}$, which implies
that adjacent energy levels repel each other. This behavior may be
attributed to the mixing between the two states related with these levels.

So far in the literature, there is no rigorous statistical description for
the transition from integrability to chaos. The nature of the stochastic
transition is more obscure in quantum than in classical mechanics, as the
assumptions that lead to the RMT description do not apply to mixed systems.
The Hamiltonian of a typical mixed system can be described as a random
matrix where some of its elements are randomly distributed and some of them
might be non-random. Moreover, the matrix elements need not all have the
same distributions and may or may not be correlated. Thus, the RMT approach
is a difficult route to follow. Comprehensive semiclassical computations
have been carried out for Hamiltonian quantum systems, which on the
classical level have a mixed phase space dynamics (see, e.g. \cite{gut} and
references therein). There have been several proposals for phenomenological
random matrix theories that interpolate between the Wigner-Dyson RMT and
banded random matrices with an almost Poissonian spectral statistics. The
standard route for the derivation is to sacrifice basis invariance but keep
matrix-element independence. The first work in this direction is due to
Rosenzweig and Porter \cite{rosen}. They model the Hamiltonian of a mixed
system by a superposition of a diagonal matrix with random elements and a
matrix drawn from a GOE. Accordingly, the variances of the diagonal elements
of the total Hamiltonian are not twice that of the off-diagonal ones, as in
the GOE case. Hussein and Pato \cite{hussein} used the maximum entropy
principle to construct such ensembles by imposing additional constraints.
Also, ensembles of banded random matrices whose entries are equal to zero
outside a band of width $b$ along the principal diagonal have been used to
model mixed systems \cite{casati,fyodorov,mirlin,kravtsov,evers}.

Another route for generalizing RMT is to conserve base invariance but allow
for the correlation of matrix elements. This has been achieved by maximizing
non-extensive entropies subject to the constraint of a fixed expectation
value of Tr$\left( H^{\dagger}H\right)$, see \cite%
{evans,toscano,nobre,abul,bertuola,abul1,abul2}. Recently, an equivalent
approach was presented in \cite{sust1,sust2}, which is based on the method
of superstatistics (statistics of a statistic) proposed by Beck and Cohen 
\cite{BC}. This formalism has been elaborated and applied successfully to a
wide variety of physical problems, e.g., in \cite%
{cohen,beck,beckL,salasnich,sattin,reynolds,ivanova,beckT}. In
thermostatics, superstatistics arises from weighted averages of ordinary
statistics (the Boltzmann factor) due to fluctuations of one or more
intensive parameters (e.g. the inverse temperature). Its application to RMT
assumes the spectrum of a mixed system as made up of many smaller cells that
are temporarily in a chaotic phase. Each cell is large enough to obey the
statistical requirements of RMT but is associated with a different
distribution of the parameter $\eta$ n Eq. (\ref{GE}) according to a
probability density $f(\eta)$. Consequently, the superstatistical
random-matrix ensemble used for the description of a mixed system consists
of a superposition of Gaussian ensembles. Its joint probability density
distribution of the matrix elements is obtained by integrating the
distribution given in Eq.~(\ref{GE}) over all positive values of $\eta$\
with a statistical weight $f(\eta)$, 
\begin{equation}
P(H)=\int_{0}^{\infty}f(\eta)\frac{\exp\left[ -\eta\text{Tr}\left(
H^{\dagger}H\right) \right] }{Z(\eta)}d\eta.  \label{PH}
\end{equation}
Despite the fact that it is hard to make this picture rigorous, there is
indeed a representation which comes close to this idea \cite{caer,muttalib}.

The present paper is concerned with a justification for the use of the
above-mentioned superstatistical generalization of RMT in the study of mixed
systems, based on the representation of their energy spectra in the form of
discrete time series in which the level order plays the role of time. The
representation of the suitably transformed eigenvalues of a quantum system
as a time series has recently allowed to determine the degree of chaoticity
of the dynamics of the system \cite%
{relano,gomez,santhanam,manimaran,santhanam1}. We have thus been motivated
by the work of Beck, Cohen and Swinney \cite{bcs} concerning the derivation
of superstatistics starting from time-series. Superstatistical thermostatics
results as a convolution of two statistics, one characterized by the
Boltzmann factor and the other corresponding to inverse-temperature
fluctuations. This requires the existence of two relaxation times. We apply
the arguments of \cite{bcs} by representing the spectra of mixed systems as
discrete time series in which the role of time is played by the level
ordering. In Section II, we consider two billiards with mushroom-shaped
boundaries as representatives of systems with mixed regular--chaotic
dynamics and three with the shape of Lima\c{c}on billiards, one of them of
chaotic and two of mixed dynamics. The quantum eigenvalues and statistical
properties of the eigenfunctions were obtained experimentally by exploiting
the equivalence of the Schr{\"o}dinger equation of a plane quantum billiard
and the Helmholtz equation for the electric field strength in a cylindrical
microwave resonator for wave lengths longer than twice the height of the
resonator. The billiards with mixed dynamics have classical phase spaces of
different structures for the two families of billards. The "time-series"
analysis of their spectra manifests the existence of two relaxation lengths,
a short one defined as the average length over which energy fluctuations are
correlated, and a long one that characterizes the typical linear size of the
heterogeneous domains of the total spectrum. It is performed in an attempt
to clarify the physical origin of the heterogeneity of the matrix-element
space, which justifies the superstatistical approach to RMT. The second main
result of this section is to derive a parameter distribution $f(\eta)$,
which is introduced in Eq.~(\ref{PH}). This paves the way for the
generalization of the Wigner surmise to superstatistics concerning the
nearest-neighbor spacing distribution (NNSD). We then apply the deduced
generalized Wigner surmise in a phenomenological analysis of the NNSD to the
measured resonance frequencies of the microwave resonators. Section IV
introduces superstatistical generalizations for the Porter-Thomas
distribution of partial widths. Then, the corresponding formulas for the
resonance strengths are used to analyze the experimental resonance-strength
distributions in the mixed Lima\c{c}on billiards. A brief summary of the
main results is given in section V.

\section{Mushroom and Lima\c{c}on Billiards}

Billiards can be used as simple models in the study of Hamiltonian systems.
They consist of a point particle which is confined to a container of some
shape and reflected elastically on impact with the boundary. The shape
determines whether the dynamics inside the billiard is regular, chaotic or
mixed. The best-known examples of chaotic billiards are the Sinai billiard
(a square table with a circular barrier at its center) and the Bunimovich
stadium (a rectangle with two circular caps) \cite{bunimovichS}. Neighboring
parallel orbits diverge when they collide with dispersing components of the
billiard boundary. In chaotic focusing billiards, neighboring parallel
orbits converge at first, but divergence prevails over convergence on
average. Divergence and convergence are balanced in integrable billiards
such as circles and ellipses.\ 

Recently Bunimovich introduced the so-called `mushroom' billiard \cite%
{bunimovichM} with the novel feature of a well-understood divided
phase-space comprising a single integrable region and a single ergodic one.
We restrict ourselves here to mushroom billiards which consist of a
semicircular region, the `hat' and a `stem', which is symmetrically attached
to its base. As the width of the stem varies from zero to the diameter of
the hat, there is a continuous transition from integrability (the semicircle
billiard) to ergodicity (in case of a rectangul stem the stadium billiard).
In mushroom billiards, the regular region has a well-defined semicircular
border about the center of the hat with radius equal to half the width of
the stem. It is composed of those trajectories in the hat that never cross
this border and therefore remain in the hat forever. Their integrability is
due to the conservation of the reflection angle for collisions with the
semicircular boundary. The chaotic component consists of trajectories that
enter the stem of the mushroom billiard. In contrast to most other mixed
systems, the dynamics of mushroom billiards is free of the usual hierarchies
of KAM islands about integrable islands in phase space. Because of its
sharply-divided phase space, mushroom billiards can be thought of as an
ideal model for the understanding of mixed dynamics. They indeed have
already been under active research \cite{altmann,tanaka,dietz,friedrich}.

The Lima\c{c}on billiard is a closed billiard whose boundary is defined by
the quadratic conformal map of the unit circle $z$ to $w$, 
\begin{equation}
w=z+\lambda z^{2}, \vert z\vert =1 .
\end{equation}

The shape of the billiard is controlled by a single parameter $\lambda$ with 
$\lambda =0$ corresponding to the circle and $\lambda =1/2$ to the cardioid
billiard \cite{backer}. For $0\leq\lambda < 1/4$, the Lima\c{c}on billiard
has a continuous and convex boundary with a strictly positive curvature and
a collection of caustics near the boundary \cite{dullin,robnik}. At $\lambda
=1/4$, the boundary has zero curvature at its point of intersection with the
negative real axis, which turns into a discontinuity for $\lambda >1/4$.
Accordingly, there the caustics no longer persist [11]. The classical
dynamics of this system and the corresponding quantum billiard have been
extensively investigated by Robnik and collaborators \cite{robnik,robnik1}.
They concluded that the dynamics in the Lima\c{c}on billiard undergoes a
smooth transition from integrable motion at $\lambda =0$ via a soft chaos
KAM regime for $0<\lambda\leq 1/4$ to a strongly chaotic dynamics for $%
\lambda =1/2$.

Both families of systems have been studied experimentally in the quantum
limit exploiting the analogy between a quantum billiard and a flat cylindric
microwave billiard \cite{rehfeld,friedrich,hesse}. The electromagnetic
resonances in a flat microwave cavity can directly be associated with
quantum states in a quantum billiard of the same geometry. For the
evaluation of statistical measures, a sufficiently large number of
resonances is needed. Experimentally, this is not trivial since each
resonance has a finite width, and as the level density increases with
frequency, single resonances can only be resolved up to a certain frequency.
Hence, to measure as many states as possible, one has to reduce the width of
the resonances by reducing the loss mechanisms. This is achieved by the use
of superconducting cavities, which are cooled down to the temperature of
liquid helium, $T=4.2$~K, in a bath cryostat. The cavities are either made
of niobium or of copper plates, which have been galvanically covered with a
layer of lead, whose thickness is several penetration depths of the
electromagnetic field.

Electric field oscillations in the interior of the cavity can be excited via
antennae. Using a vectorial network analyzer, the complex amplitude ratio of
input to output signal from the cavity can be measured. Peaks in the modulus
of the amplitude are found at resonance frequencies, corresponding to
eigenmodes of the system. There, the wave field forms a standing wave. It
can be only excited when the antennae used in the measurement process are
not near nodal lines of the wave function, where the amplitude is very
small. Thus, usually several antennae are used to measure different spectra,
enabling to identify \textit{all} resonances up to a certain frequency. For
flat resonators of a length scale of 30 cm, one collects approximately the
first 800 eigenvalues with a very high precision. Experimental data of both
kinds of systems considered in this paper have been obtained via this
procedure \cite{richter_ima}.

\begin{figure}[ptb]
\begin{center}
\includegraphics[width=\columnwidth]{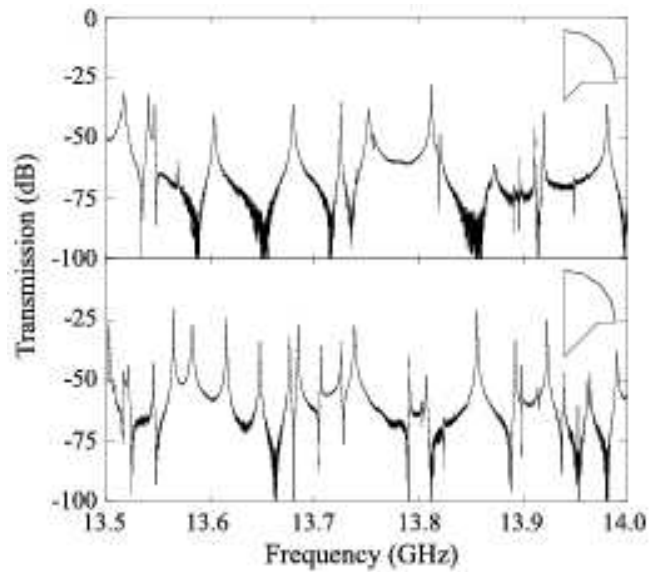}
\end{center}
\caption{Part of the transmission spectrum of the small (left panel) and
large (right panel) mushroom billiard. The insets show the geometry of the
billiards.}
\label{spectr}
\end{figure}

Two mushroom billiards have been recently investigated experimentally. In
order to avoid symmetry effects and bouncing ball orbits between parallel
walls in the stem, their shape is of a half mushroom with a slant stem (see
inset of Fig.~\ref{spectr}. The ratio of the width of the stem to the
diameter of the hat is 1:3 (2:3) for the small (large) mushroom billiard.
The degree of chaos, which is the measure of all chaotic parts of the phase
space, is 45.5~$\%$ (82.9~$\%$), and the first 780 (938) resonances could be
detected. 
In Fig.~\ref{spectr} we show a part of the spectra of the small and large
mushroom billiard, respectively. Details of the experiment with the larger
mushroom billiard can be found in \cite{friedrich}. Three desymmetrized
cavities with the shape of billiards from the family of Lima\c{c}on
billiards have been constructed for the values $\lambda={0.125, 0.150, 0.300}
$ and the first 1163, 1173 and 942 eigenvalues were measured, respectively.
More details on these experiments are given in \cite{rehfeld, hesse}. To
compare the statistical properties of the eigenvalues with universal
predictions considered in the present paper, they have to be rescaled to
unit mean spacing. This is done by an unfolding procedure using Weyl's
formula \cite{Weyl}, which relates the billiard area and circumference to
the number of resonance frequencies below a given one.

As outlined above, the phase space structure of the billiards with mixed
dynamics are different for the two families under consideration. The
experimental NNSD of the larger of the two mushroom billiards exhibits \cite%
{friedrich} a statistically significant dip at $s\approx0.7$. It vanishes
when the contribution of the two shortest periodic orbits is subtracted.
Such a dip has never been observed in the spectra of other billiards with
mixed dynamics, including the Lima\c{c}on billiards considered here. We
shall show below that nevertheless the statistical properties of the Lima%
\c{c}on billiards are indistinguishable from those of the mushroom billiards
after removal of the contribution of these periodic orbits, in spite of the
difference of their phase space structure.

\section{Time-series representation}

In this section the time series method used for the study of the
fluctuations of the resonance spectra of the mushroom billiards is
introduced. Representing energy levels of a quantum system as a discrete
time series has been probed in a number of recent publications. Rela\~{n}o
et al. \cite{relano} considered a sequence of energy levels as a discrete
time series in which the energy played the role of time. They conjectured
that the power spectra of chaotic quantum systems are characterized by $1/f$
noise, whereas integrable quantum systems exhibit $1/f^{2}$ noise. This
conjecture was supported by numerical experiments which involved classical
random-matrix ensembles and atomic nuclei. Moreover, the power spectrum of
an experimentally modeled quantum Sinai billiard exhibits a clear $1/f$
noise through almost the whole frequency domain. Mixed systems on the other
hand, like the Lima\c{c}on billiard, the quartic coupled oscillator, and the
kicked top, are characterized by a $1/f^{\alpha}$ noise \cite%
{gomez,santhanam}. In all these cases, the exponent $\alpha$ was related to
the degree of chaos. Manimaran et al. \cite{manimaran} recently developed a\
wavelet based approach to discrete time series and employed it to
characterize the scaling behavior of spectral fluctuations of random matrix
ensembles, as well as complex atomic systems. Santhanam et al. \cite%
{santhanam1} studied the spectra of atoms and Gaussian ensembles using the
detrended fluctuation analysis, which is a popular tool to study long range
correlations in time series \cite{peng}. They showed that this analysis is
related to the $\Delta_{3}$ statistics of RMT.

Here we apply the time-series analysis to the study of the energy spectra of
mixed systems from another point of view. Our goal is to test the hypothesis
that quantum systems can be modeled by a generalization of RMT \cite%
{sust1,sust2} based on the concept of superstatistics \cite{BC}. We follow
an approach recently proposed by Beck, Cohen and Swinney \cite{bcs}, which
describes how to proceed from a given experimental time series to a
superstatistical description. This approach allows one to check whether a
time series contains two separate time scales, and also to extract the
relevant probability densities of superstatistical parameters from the time
series.

\subsection{Spectral relaxation lengths}

Superstatistical RMT assumes that the space of matrix elements consists of
many spatial cells with different values of some intensive parameter, e.g.
the inverse variance $\eta$. In systems with mixed regular-chaotic dynamics,
the origin of this spectral heterogeneity is the possible partial
conservation of an unknown or ignored symmetry. In the space of matrix
elements each heterogeneous domain comprises those matrix elements that
couple states, which have similar properties with respect to this symmetry,
where the typical size of the heterogeneous cells $T$ measures the
correlation length in that space. The heterogeneity of the space of matrix
elements presumably causes one in the structure of the spectrum. Each cell
is assumed to reach local equilibrium very fast, i.e., the associated
relaxation length $\tau$, which is defined as that length-scale over which
energy fluctuations are correlated, is short. It may also be regarded as an
operational definition for the average energy separation between levels due
to level repulsion. In the long-term run, the stationary distributions of
this inhomogeneous system arise as a superposition of the "Boltzmann
factors" of the standard RMT, i.e. $e^{-\eta\text{Tr}H^{2}}$. The parameter $%
\eta$ is approximately constant in each cell for an eigenvalue interval of
length $T$. In superstatistics this superposition is performed by weighting
the stationary distribution of each cell with the probability density $%
f(\eta)$ to observe some value $\eta$ in a randomly chosen cell and
integrating over $\eta$. Of course, a necessary condition for a
superstatistical description to make sense is the condition $\tau\ll T$,
because otherwise the system is not able to reach local equilibrium before
the next change takes place.

Our goal is to show that the behavior of a fictitious time series formed by
the 780 and 938 resonances, respectively, in the two mushroom billiards, and
the approximately 1100 resonances in each of the three Lima\c{c}on billiards
is consistent with superstatistics. For this purpose the distribution $%
f(\eta)$ is derived by proceeding as in \cite{bcs}. We extract the
relaxation lengths (times) to local equilibrium $\tau$ and the large length
scale $T$ on which the intensive parameter fluctuates and show that there is
a clear scale separation of the spectral correlations in each billiard.

\begin{figure}[ptb]
\begin{center}
\includegraphics[width=\columnwidth,height=5cm]{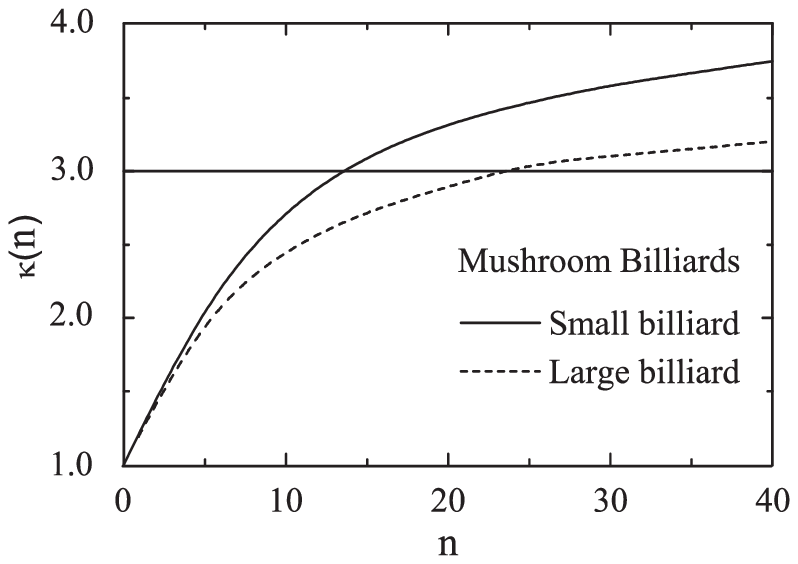} %
\includegraphics[width=\columnwidth,height=5cm]{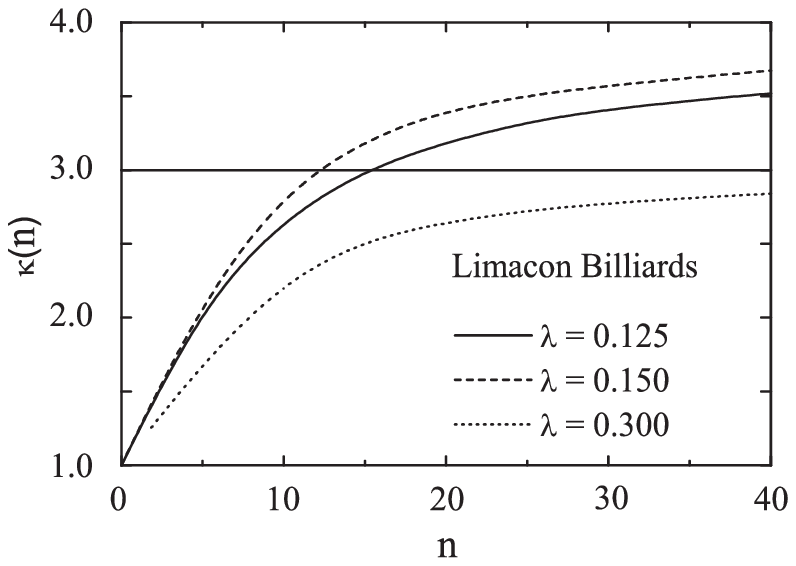}
\end{center}
\caption{ The local kurtosis of the spacing intervals of the mushroom (left
panel) and Lima\c{c}on (right panel) billiards.}
\label{fig_1}
\end{figure}

\begin{figure}[ptb]
\begin{center}
\includegraphics[width=\columnwidth,height=5cm]{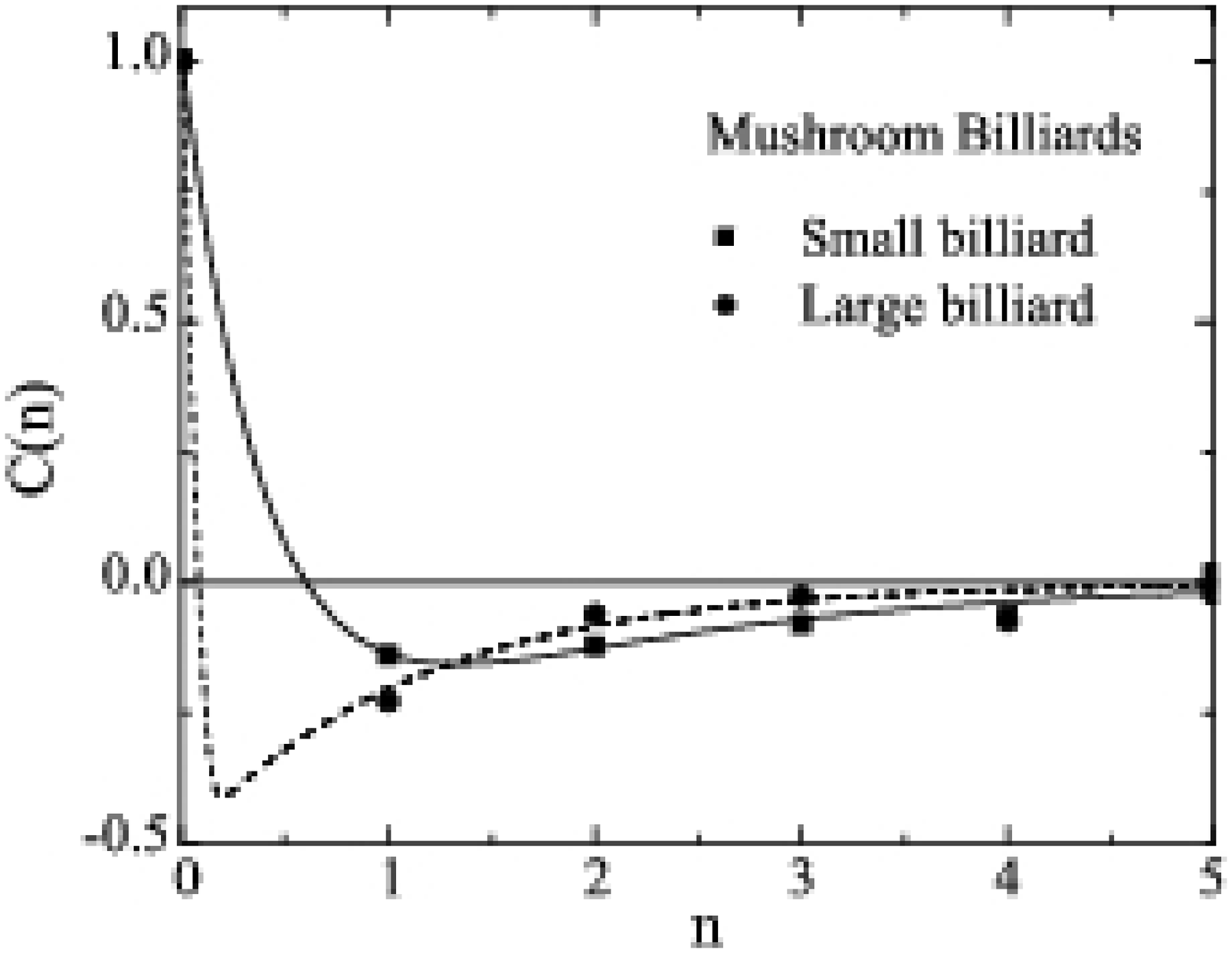} %
\includegraphics[width=\columnwidth,height=5cm]{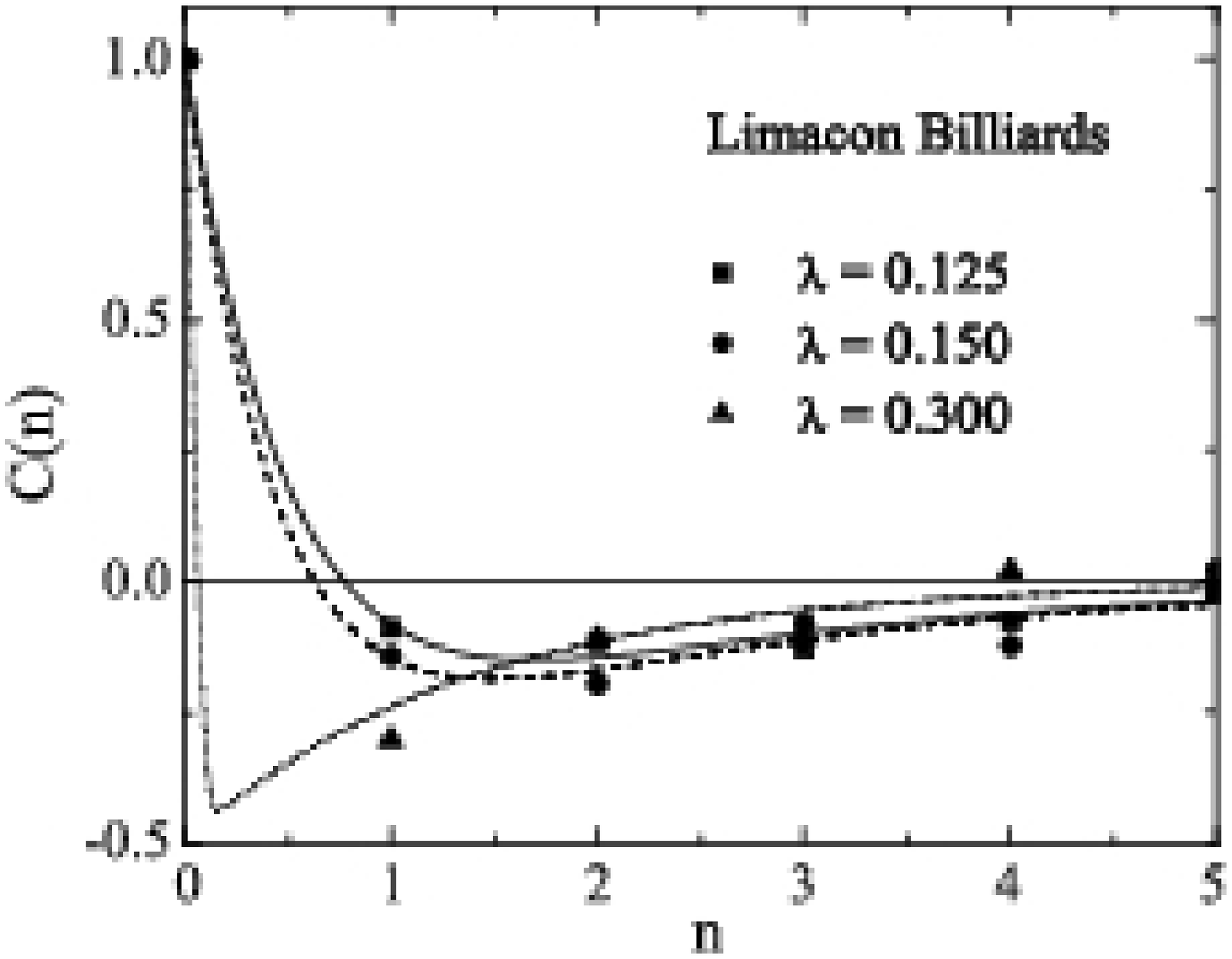}
\end{center}
\caption{ The autocorrelation of the spacings within spectral intervals for
the mushroom (left panel) and Lima\c{c}on (right panel) billiards.}
\label{fig_2}
\end{figure}

First, let us determine the long time scale $T$. For this we divide the
spacings series into $N$ equal level-number intervals of size $n$. The total
length of the spectrum is $N\cdot n$. We then define the mean local kurtosis 
$\kappa(n)$ of a spacing interval of length $n$ by%
\begin{equation}
\kappa(n)=\frac{1}{N}\sum_{i=1}^{N}\frac{\left\langle (s-\overline{s}%
)^{4}\right\rangle _{i,N}}{\left\langle (s-\overline{s})^{2}\right\rangle
_{i,N}^{2}}.  \label{kappa}
\end{equation}
Here $\left\langle \cdots\right\rangle _{i,N}=\sum_{k=(i-1)\cdot
n+1}^{i\cdot n}\cdots$ denotes a summation over an interval of length $n$
starting at level spacing $in$, and $\overline {s}$ is either the local
average spacing in each spacing interval or the global average $\overline{s}%
=1$ over the entire spacings series. We chose the latter one. In probability
theory and statistics, kurtosis is a measure for the "flatness" of the
probability distribution of a real-valued random variable. Higher kurtosis
means that a larger part of the contributions to the variance is due to
infrequent extreme deviations, as opposed to frequent modestly sized ones. A
superposition of local Gaussians with local flatness three results in a
kurtosis of three. We define the superstatistical level-number scale $T$ by
the condition%
\begin{equation}
\kappa(T)=3,  \label{kappaT}
\end{equation}
that is, we look for the simplest superstatistics, a superposition of local
Gaussians \cite{bcs}. If $n$ is chosen such, that only one value of $s$ is
contained in each interval, then of course $\kappa(1)$ =1. If on the other
hand $n$ comprises the entire spacing series, then we obtain the flatness of
the distribution of the entire signal, which will be larger than 3, since
superstatistical distributions are fat-tailed. Therefore, there exists a
level-number scale $T$ which solves Eq.~(\ref{kappaT}). Figure~\ref{fig_1}
shows the dependence of the local flatness of a spacing interval on its
length for the two mushroom and the three Lima\c{c}on billiards. 
In the case of the chaotic Lima\c{c}on billiard, in which $\lambda=0.300,$
the quantity $\kappa$ does not cross the line of $\kappa=3$ for the
considered values of $n$. It is expected that $T=N$ in this case since the
fluctuations in a chaotic (unfolded) spectrum are uniform. The values of $T$
for the mixed billiards are given in Table~\ref{table1}. 
\begin{table}[tbp]
\caption{Correlation lengths estimated from the time series representing the
frequency spectra of the mushroom and Lima\c{c}on billiards.}
\label{table1}%
\begin{centering}
%\begin{tabular*}{\linewidth}{c@{\extracolsep\fill}ccccc}
\begin{tabular*}{\columnwidth}{c@{\extracolsep\fill}ccccc}
\hline\hline & \multicolumn{2}{c}{Mushroom} &
\multicolumn{3}{c}{Lima\c{c}on}\\
& small & large & $\lambda=0.125$ & $\lambda=0.150$ &
$\lambda=0.300$\\
\hline
\multicolumn{6}{c}{Long-range correlation time}\\
$T$ & 12.2 & 23.5 & 12.3 & 15.4 & large\\
\multicolumn{6}{c}{Short-range correlation times}\\
$\tau_{1}$ & 0.40 & 0.03 & 0.51 & 0.44 & 0.02\\
$\tau_{2}$ & 1.55 & 1.12 & 1.9 & 2.01 & 1.36\\
\hline\hline
\end{tabular*}
\end{centering}
\end{table}
The short time scale, that is the relaxation time associated with each of
the $N$ intervals, was estimated in \cite{bcs} from the small-argument
exponential decay of the autocorrelation function%
\begin{equation}
C_{s}(n)=\frac{\overline{s(i)s(i+n)}-1}{\overline{s^{2}}-1}
\end{equation}
of the time series $s(t)$ under consideration. Figure~\ref{fig_2} shows the
behavior of the autocorrelation functions for the series of
resonance-spacings of the two families of billiards. 
Quite frequently, the autocorrelation function shows single-exponential
decays,$\ C(n)=e^{-n/\tau}$, where $\tau>0$ defines a relaxation "time". A
typical example is the velocity correlation of Brownian motion \cite{reif}.
The autocorrelation functions studied here clearly do not follow this trend.
For the systems with mixed dynamics, they decay rapidly from a value of $%
C(0)=1$, change sign at some $n\sim1$ becoming negative, then asymptotically
tend to zero. In an attempt to quantify the dependence of $C_{s}$ on $n$, we
parametrized its empirical value in the form of a superposition of two
exponentially decaying functions%
\begin{equation}
C_{s}(n)=A_{1}e^{-n/\tau_{1}}+A_{2}e^{-n/\tau_{2}}
\end{equation}
and (arbitrarily) fixed the superposition coefficient as $A_{1}=1.5$ and $%
A_{2}=-0.5.$ The curves in Fig.~\ref{fig_2} show the resulting
parametrization. The best fit parameters are given in Table~\ref{table1}. We
may estimate $\tau$ as the mean values of $\tau_{1}$ and $\tau_{2}$\ and
conclude that $\tau$ has a value slightly larger than 1 for each billiard.
This is sufficient to conclude that the ratio $T/\tau$ is large enough in
each billiard to claim two well separated "time" scales in the
level-spacings series, which justifies describing them within the framework
of superstatistics.

Both Figs.~\ref{fig_1} and \ref{fig_2} as well as Table~\label{table1}
suggest that the evolution of the quantities $\kappa(n)$ and $C_{s}(n)$
along the time series for the two billiard families is the same despite the
different behavior of their classical dynamics.

\subsection{Estimation of the parameter distribution}

The distribution $f(\eta)$ is determined by the spatiotemporal dynamics of
the entire system under consideration. Beck et al. \cite{bcs} have argued
that typical experimental data are described by one of three
superstatistical universality classes, namely, $\chi^{2}$, inverse $\chi^{2}$%
, or log-normal superstatistics. The first is the appropriate one if $\eta$
has contributions from $\nu$ Gaussian random variables $X_{1}$, . . . , $%
X_{\nu}$ due to various relevant degrees of freedom in the system. As
mentioned before $\eta$ needs to be positive; this is achieved by squaring
these Gaussian random variables. Hence, $\eta=\sum_{i=1}^{\nu}X_{i}^{2}$ is $%
\chi^{2}$ distributed with degree $\nu$,%
\begin{equation}
f(\eta)=\frac{1}{\Gamma(\nu/2)}\left( \frac{\nu}{2\eta_{0}}\right) ^{\nu
/2}\eta^{\nu/2-1}e^{-\nu\eta/2\eta_{0}}.  \label{P1}
\end{equation}
The average of $\eta$ is $\eta_0=\int_0^\infty\eta f(\eta )d\eta $. The same
considerations are applicable if\ $\eta^{-1},$ rather than $\eta,$ is the
sum of several squared Gaussian random variables. The resulting distribution 
$f(\eta)$ is the inverse $\chi^{2}$ distribution given by%
\begin{equation}
f(\eta)=\frac{\eta_{0}}{\Gamma(\nu/2)}\left( \frac{\nu\eta_{0}}{2}\right)
^{\nu/2}\eta^{-\nu/2-2}e^{-\nu\eta_{0}/2\eta},  \label{P2}
\end{equation}
where again $\eta_0$ is the average of $\eta$. Instead of being a sum of
many contributions, the random variable $\eta$ may be generated by
multiplicative random processes. Then $\ln\eta=\sum_{i=1}^{\nu}\ln X_{i}$ is
a sum of Gaussian random variables. Thus it is log-normally distributed, 
\begin{equation}
f(\eta)=\frac{1}{\sqrt{2\pi}v\eta}e^{-\left. \left[ \ln(\eta/\mu)\right]
^{2}\right/ 2v^{2}},  \label{P3}
\end{equation}
which has an average $\mu\sqrt{w}$ and variance $\mu^{2}w(w-1)$, where $%
w=\exp(v^{2})$.

\begin{figure}[ptb]
\begin{center}
\includegraphics[width=\columnwidth,height=5cm]{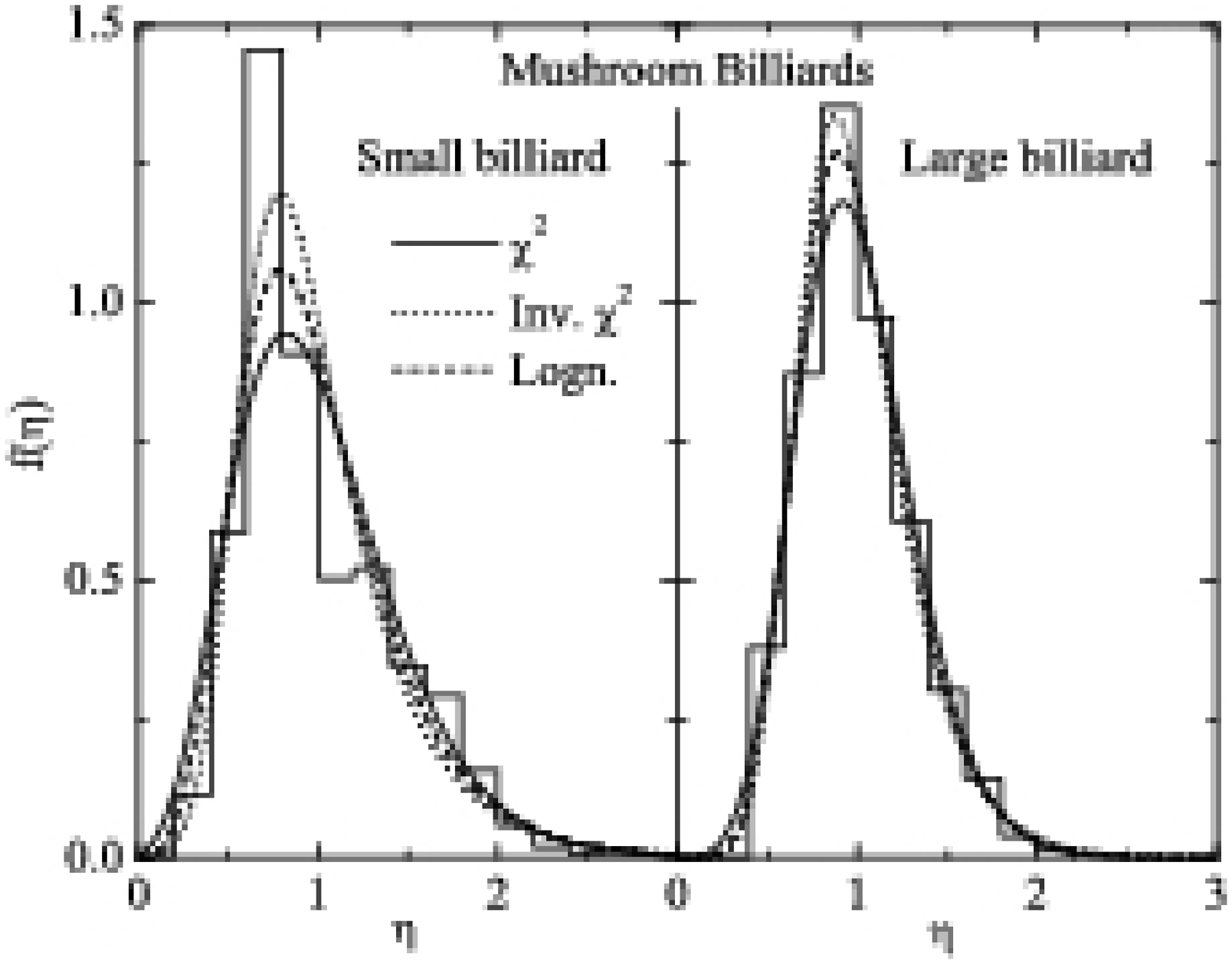} %
\includegraphics[width=\columnwidth,height=5cm]{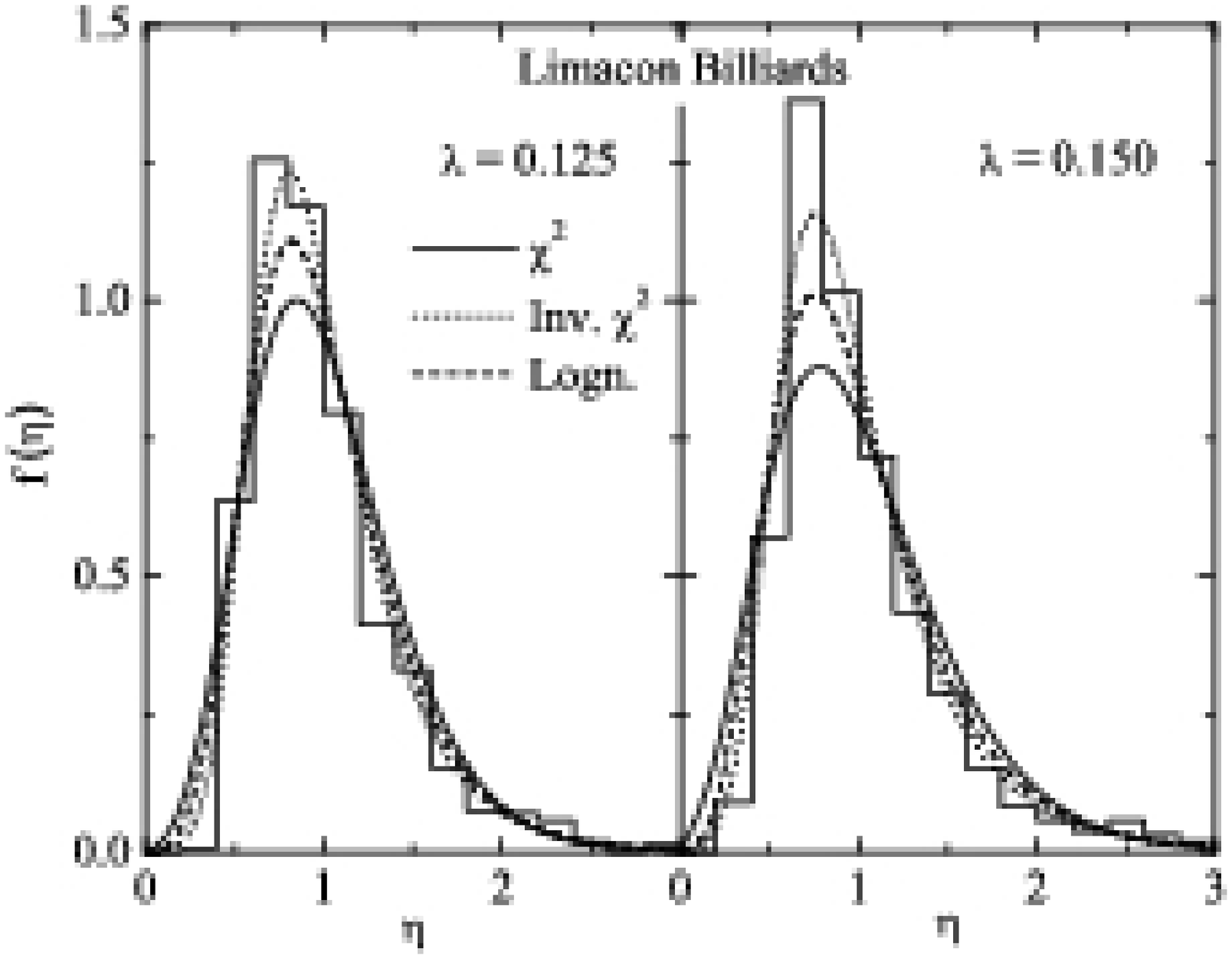}
\end{center}
\caption{Estimation of the parameter distribution for the superstatistical
description of spectra of the mushroom (left panel) and Lima\c{c}on (right
panel) billiards. The solid lines represent the $\protect\chi^{2}$, the
daotted lines the inverse $\protect\chi^{2}$ and the dashed lines the
log-normal distribution.}
\label{fig_3}
\end{figure}

Next, we need to determine which of these distributions fits best that of
the slowly varying stochastic process $\eta(t)$ described by the
experimental data. Since the variance of superimposed local Gaussians (see
remark after Eq.~(\ref{kappaT})) is given by $\eta^{-1}$, we may determine
the process $\eta(t)$ from the series%
\begin{equation}
\eta(i)=\frac{1}{\left\langle s^{2}\right\rangle _{i,T}-\left\langle
s\right\rangle _{i,T}^{2}}.
\end{equation}
Accordingly, the probability density $f(\eta)$ is determined from the
histogram of the $\eta\left( i\right)$ values for all $i$; the resulting
experimental distributions are shown in Fig.~\ref{fig_3}. 
We compared them with the log-normal, the $\chi^{2}$ and the inverse $%
\chi^{2}$ distributions with the same mean $\left\langle \eta\right\rangle $
and variance $\left\langle \eta^{2}\right\rangle -\left\langle
\eta\right\rangle ^{2}$. The inverse $\chi^{2}$ distribution fits the data
significantly better than the other two distributions.

\section{Nearest-neighbor spacing distribution}

This section focusses on the question whether the inverse $\chi ^{2}$
distribution of the superstatistical parameter $\eta$ in Eq.~(\ref{P2}) is
suitable for describing the NNSD of systems in the transition out of chaos
within the superstatistical approach to RMT. As mentioned above, the NNSD of
a chaotic system is well described by that of random matrices from the GOE,
the Wigner surmise, if the system is chaotic and by that for Poisson
statistics if it is integrable. Numerous interpolation formulas describing
the intermediate situation between integrability and chaos have been
proposed \cite{guhr}. One of the most popular ones is that introduced by
Brody \cite{brody} although it is purely phenomenological. This distribution
coincides with the Wigner distribution for a fully chaotic and with
Poisson's for an integrable system. It is known to provide an excellent
description for the NNSDs of numerous mixed systems. Another
phenomenological distribution was proposed in \cite{casati} and its
usefulness was demonstrated for band random matrices. Lenz and Haake \cite%
{lenz} derived a distribution based on the model of additive random
matrices. Finally, Berry and Robnik elaborated a NNSD for mixed systems
based on the assumption that semiclassically the eigenfunctions and
associated Wigner distributions are localized either in classically regular
or chaotic regions in phase space \cite{berryrobnik}. Accordingly, the
sequences of eigenvalues connected with these regions are assumed to be
statistically independent, and their mean spacing is determined by the
invariant measure of the corresponding regions in phase space. The largest
discrepancy between the Brody and the Berry-Robnik distribution is observed
for level spacings $s$ close to zero. While the former vanishes for $s=0$,
the latter approaches a constant and nonvanishing value for $s\to 0$. In 
\cite{friedrich,rehfeld} the experimental NNSDs for the measured resonance
frequencies of the mushroom and the Lima\c{c}on billiards, respectively,
were compared to the Berry-Robnik (BR) distribution.

It follows from Eq.~(\ref{PH}) that the statistical measures of the
eigenvalues of the superstatistical ensemble are obtained as an average of
the corresponding $\eta$-dependent ones of standard RMT weighted with the
parameter distribution $f(\eta)$. In particular, the superstatistical NNSD
is given by \cite{sust1} as 
\begin{equation}
p(s)=\int_{0}^{\infty}f(\eta)~p_{\text{W}}(\eta,s)~d\eta,
\end{equation}
where $p_{\text{W}}(\eta,s)$ is the Wigner surmise for the Gaussian
orthogonal ensemble with the mean spacing depending on the parameter $\eta$,%
\begin{equation}
p_{\text{W}}(\eta,s)=\eta s\exp\left( -\frac{1}{2}\eta s^{2}\right) .
\label{PS}
\end{equation}
For a $\chi^{2}$ distribution of the superstatistical parameter $\eta$, one
substitutes Eq.~(\ref{P1}) into Eq.~(\ref{PS}) and integrates over $\eta.$
The resulting NNSD is given by 
\begin{equation}
p_{\chi^2}(\nu,s)=\frac{\eta_{0}s}{(1+ \eta_{0}s^{2}/\nu)^{1+\nu/2}}.
\label{PS1}
\end{equation}
The parameter $\eta_{0}$ is fixed by requiring that the mean-level spacing $%
\left\langle s\right\rangle $ equals unity, yielding 
\begin{equation}
\eta_{0}=\frac{\pi\nu}{4}\left[ \left. \Gamma\left( \frac{\nu-1}{2}\right)
\right/ \Gamma\left( \frac{\nu}{2}\right) \right] ^{2}.
\end{equation}
For an inverse $\chi^{2}$ distribution of $\eta$, given by Eq.~(\ref{P2}),
one obtains the following superstatistical NNSD, 
\begin{equation}
p_{\,\text{Inv}\,\chi^{2}}(\nu,s)=\frac{2\eta_0s}{\Gamma\left(\nu/2\right)}%
\left( \sqrt{\eta_0\nu}s/2\right)^{\nu/2} \\
K_{\nu/2}\left(\sqrt{\eta_0\nu}s\right),  \label{PS2}
\end{equation}
where $K_{m}(x)$ is a modified Bessel function \cite{grad} and $\eta_0$
again is determined by the requirement that the mean-level spacing $%
\left\langle s\right\rangle $ equals unity, 
\begin{equation}
\eta_0=\frac{4\pi}{\nu^{3}}\left[\left.\Gamma\left(\frac{3+\nu}{2}\right)
\right/ \Gamma\left(\frac{\nu}{2}\right)\right] ^{2}.
\end{equation}
Finally, if the parameter $\eta~$has a normal distribution (10), then the
NNSD 
\begin{equation}
p_{\text{\,logn}}(v,s)=\frac{s}{\sqrt{2\pi}v}\int_{0}^{\infty}\exp\left[ -%
\frac{\eta s^{2}}{2}-\frac{\ln^{2}\left( \frac{2}{\pi}\eta
e^{-v^{2}/4}\right) }{2v^{2}}\right] d\eta  \label{PS3}
\end{equation}
can only be evaluated numerically.

\begin{figure}[ptb]
\begin{center}
\includegraphics[width=\columnwidth,height=5cm]{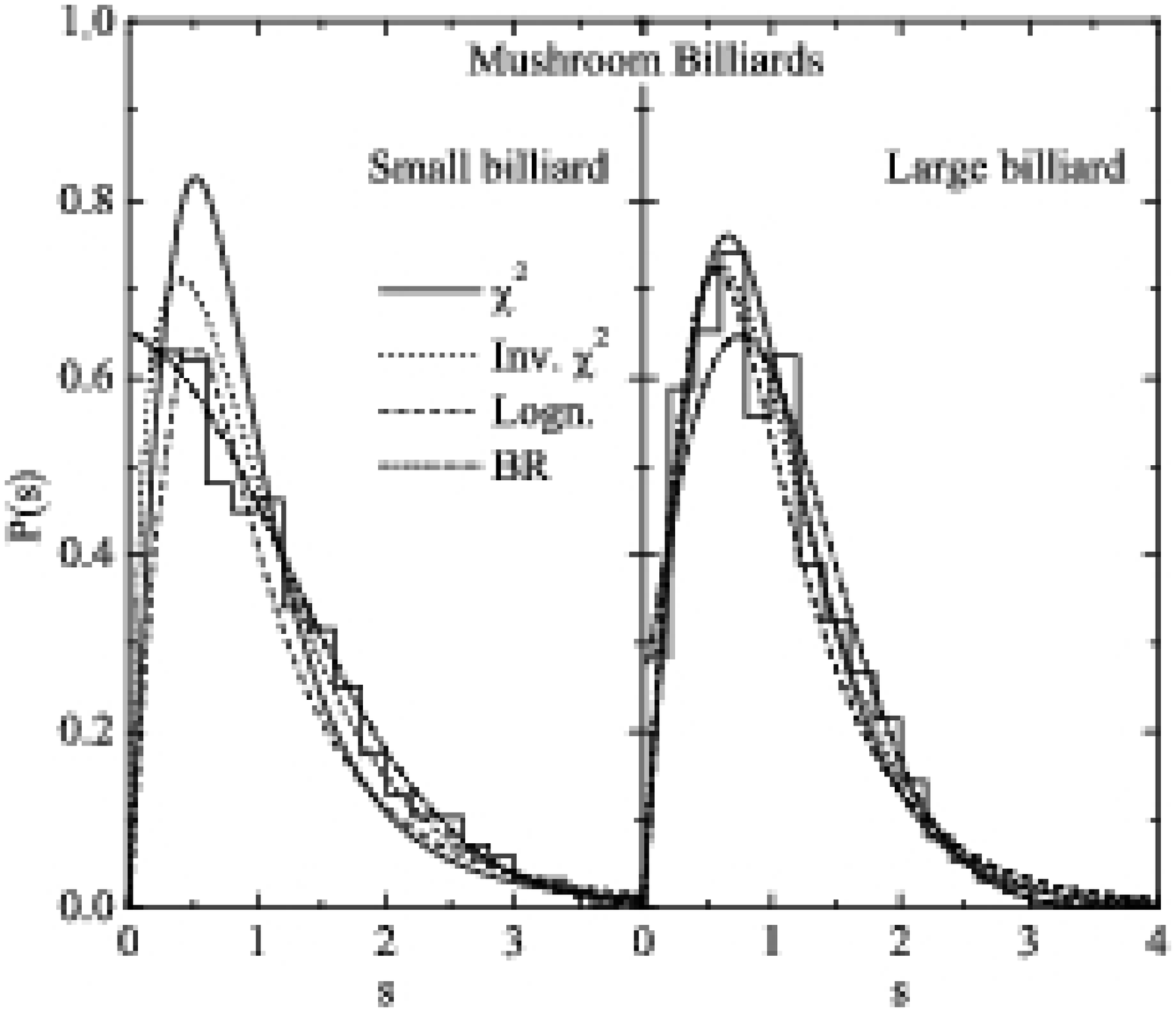} %
\includegraphics[width=\columnwidth,height=5cm]{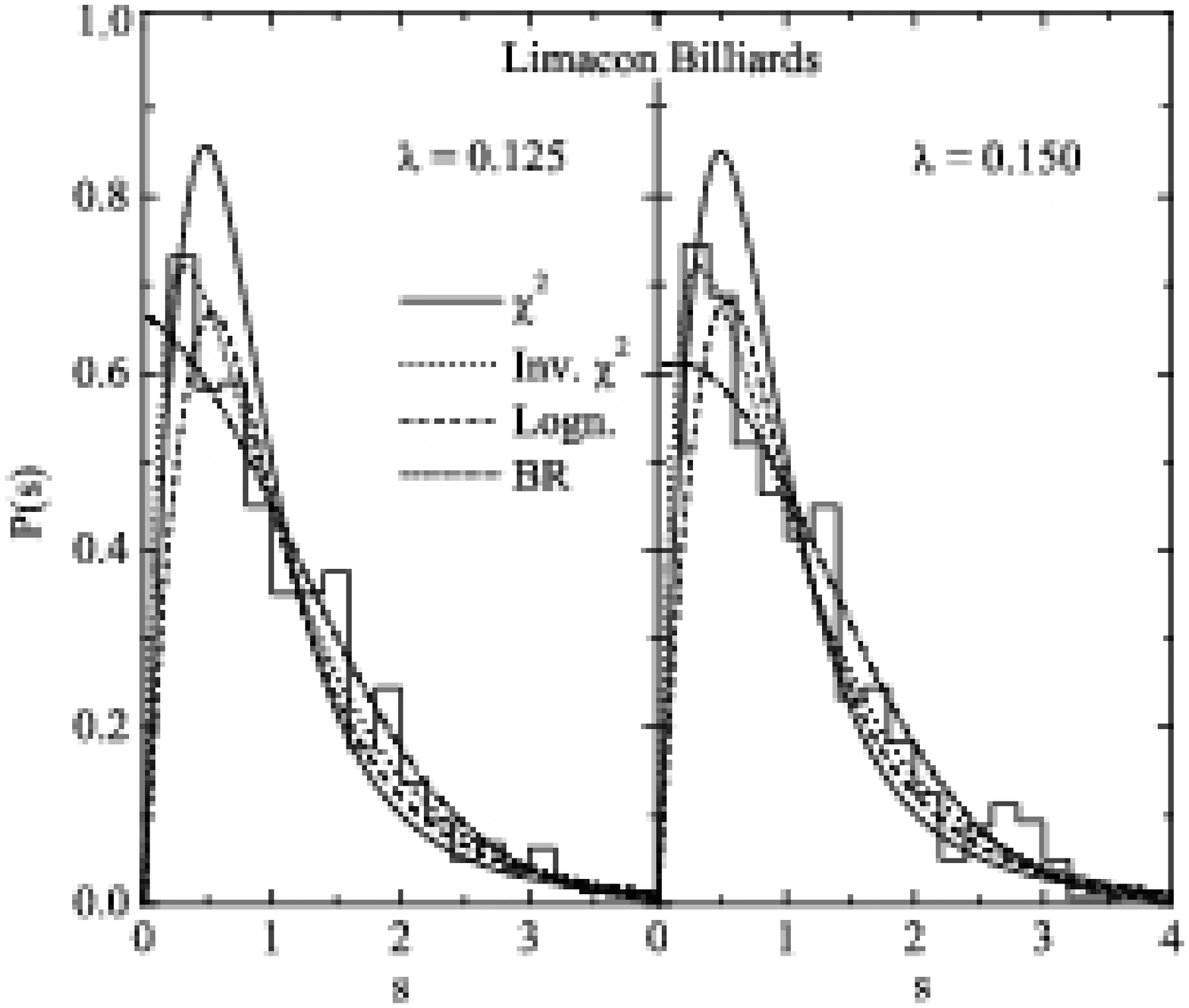}
\end{center}
\caption{ Experimental NNSDs for the mushroom (left panel) and Lima\c{c}on
(right panel) billiards compared with the superstatistical distributions.
The solid lines represent the $\protect\chi^{2}$, the dashed lines the
inverse $\protect\chi^{2}$, the dashed lines the log-normal and the
short-dashed lines the Berry-Robnik distribution.}
\label{fig_4}
\end{figure}

We compared the resulting NNSDs given in Eqs.~(\ref{PS1}), (\ref{PS2}) and (%
\ref{PS3}) with the experimental ones for the mushroom billiards and the two
Lima\c{c}on billiards with mixed dynamics. In Fig.~\ref{fig_4} the
experimental results are shown together with the superstatistical and the BR
distributions. 
The best fit values of the parameters are given in Table~\ref{table2}. 
\begin{table*}[tbp]
\caption{Best-fit parameters for the experimental NNSD and
resonance-strength distributions in the mixed billiards. The corresponding $%
\protect\chi^{2}$-values are given in brackets.}
\label{table2}%
\begin{centering}
%\begin{tabular*}{\linewidth}{l@{\extracolsep\fill}cccc}
\begin{tabular*}{\linewidth}{l@{\extracolsep\fill}cccc}
\hline\hline & \multicolumn{2}{c}{Mushroom} &
\multicolumn{2}{c}{Lima\c{c}on}\\& small & large & $\lambda=0.125$ & $\lambda=0.150$\\
\hline \multicolumn{5}{l}{NNSD}\\ 
$\chi^{2}$ & $\nu=2.95$ (0.0189)
& $\nu=6.03$ (0.0031) & $\nu=2.57$ (0.0133) & $\nu=2.63$ (0.0097)\\ 
Inverse-$\chi^{2}$ & $\nu=0.00$ (0.0042) 
& $\nu=2.31$ (0.0018) & $\nu=0.00$ (0.0031) & $\nu=0.00$ (0.0021)\\ 
Log-normal & $v=1.41$ (0.0129) 
& $v=0.96$ (0.0033) & $v=1.23$ (0.0104) & $v=1.17$ (0.0078)\\
Berry-Robnik& $q=0.24$ (0.0007) & $q=0.87$ (0.0027) & $q=0.58$
(0.0020) & $q=0.62$ (0.0034)\\ 
\\
\multicolumn{5}{l}{Resonance strength distribution}\\ 
\multicolumn{3}{l}{$\chi^{2}$} &
$\nu=2.66$ (0.00142) & $\nu=3.47$ (0.00035)\\
\multicolumn{3}{l}{Inverse-$\chi^{2}$} & $\nu=0.50$ (0.00030) &
$\nu=0.98$ (0.00045)\\ 
\multicolumn{3}{l}{Gamma distribution} &
$\nu=0.76$ (0.00120) & $\nu=0.81$ (0.00131)\\\hline\hline
\end{tabular*}
\end{centering}
\end{table*}
We in addition used the entropy \cite{Kullback,barbosa} 
\begin{equation}
S(P\left\vert p\right) =-\int P(x)\ln\frac{P(x)}{p(x)}dx
\end{equation}
to express the difference between the theoretical Ansatz $p(x)$ and the
experimental distribution $P(x)$. Such a reference term has been discussed
in the literature in the context of going from a discrete to a continuous
system and is proportional to the limiting density of discrete points \cite%
{Jaynes}. As is well known, the entropy $S(P\left\vert p\right)$ is negative
everywhere except at its maximum, where it equals zero and $P(x)=p(x)$. One
can use this fact to find the value of the parameter of a theoretical
distribution that has the least distance to the experimental distribution.
The resulting best fit values of the parameters are given in Table~\ref%
{table3}. 
\begin{table*}[tbp]
\caption{Best-fit parameters obtained by extremizing the relative entropy
for the experimental NNSDs the mixed billiards. The corresponding values of
the relative entropy are given in brackets.}
\label{table3}%
\begin{centering}
%\begin{tabular*}{\linewidth}{l@{\extracolsep\fill}cccc}
\begin{tabular*}{\linewidth}{l@{\extracolsep\fill}cccc}
\hline\hline & \multicolumn{2}{c}{Mushroom} &
\multicolumn{2}{c}{Lima\c{c}on}\\ & small & large &
$\lambda=0.125$ & $\lambda=0.150$\\\hline
\multicolumn{5}{l}{NNSD}\\ $\chi^{2}$ & $\nu=2.67$ (-0.0920) &
$\nu=7.93$ (-0.0232) & $\nu=2.62$ (-0.0909) & $\nu=2.69$ (-0.0827)\\
Inverse-$\chi^{2}$ & $\nu=0.01$ (-0.0353) & $\nu=3.14$
(-0.0175) & $\nu=0.01$ (-0.0341) & $\nu=0.01$ 
(-0.0362)\\ Log-normal & $v=1.31$ (-0.0568) & $v=0.60$ (-0.0206) &
$v=1.33$ (-0.0533) & $v=1.26$ (-0.0537)\\ Berry-Robnik & $q=0.55$
(-0.0289) & $q=0.87$ (-0.0203) & $q=0.57$ (-0.0293) & $q=0.58$
(-0.0378)\\\hline\hline
\end{tabular*}
\end{centering}
\end{table*}
Figure~\ref{fig_4} and Tables~\ref{table2} and \ref{table3} suggest the
validity of the superstatistical distribution, especially for the nearly
chaotic billiards. It clearly shows that the NNSD for the inverse $\chi^{2}$
distribution $p_{\,\text{Inv}\,\chi^{2}}(\nu,s)$ agrees in a similar or,
especially for small spacings, even better quality with the experimental
ones as the others including the Brody (not shown) and the BR distribution.

\begin{figure}[ptb]
\begin{center}
\includegraphics[width=\columnwidth,height=5cm]{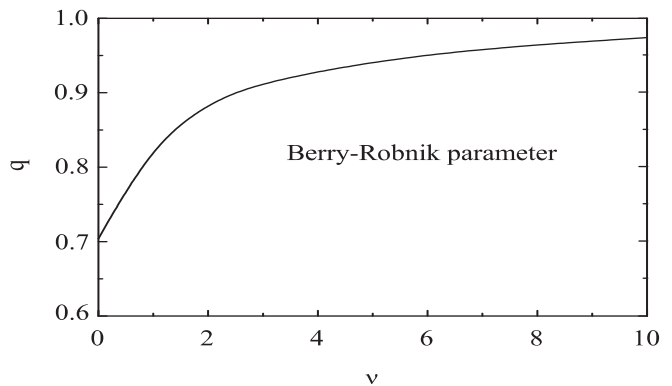}
\end{center}
\caption{ The relation between the tuning parameters $q$ and $\protect\nu$
of the Berry-Robnik and superstatistical NNSDs, respectively.}
\label{fig_5}
\end{figure}

The distribution $p_{\,\text{Inv}\,\chi^{2}}(\nu,s)$ (see Eq.~(\ref{PS2}))
coincides with the Wigner distribution in the limit of $\nu\rightarrow\infty.
$ As $\nu$ decreases, the distribution evolves towards a well-defined limit,
but this limiting case does not resemble the Poisson distribution as one
would expect. To demonstrate this behavior and give a feeling for the size
of the tuning parameter $\nu,$\ we evaluate its value corresponding to the
minimal deviation from a BR distribution with a given degree of chaoticity $q
$. For this we define a measure%
\begin{equation}
d_{\,\text{Inv}\,\chi^{2},\text{BR}}(\nu,q)=\min{\int_{0}^{\infty}\left[
p_{\,\text{Inv}\,\chi^{2}}(\nu,s)-p_{\text{BR}}(q,s)\right] ^{2}ds}.
\end{equation}
The distance $d_{\,\text{Inv}\,\chi^{2},\text{BR}}$ between the two
distributions equals zero for $q=1$ and $\nu\rightarrow\infty$, where both
distributions coincide with the Wigner distribution. Its value increases on
departure from these parameter values, has a maximum value of 0.0030 for $%
\nu=1$ and then decreases to a value of 0.0025 at $\nu=0$. The measure $d_{\,%
\text{Inv}\,\chi^{2},\text{BR}}(\nu,q)$\ yields a relation between the BR
and the superstatistical parameter $\nu$, which is shown in Fig.~\ref{fig_5}%
. 
The figure suggests that $p_{\,\text{Inv}\,\chi^{2}}(\nu,s)$ can only
describe the initial stage of the transition from chaos to regularity. One
can find a BR distribution that agrees well with $p_{\,\text{Inv}\,\chi^{2}}$
until an intermediate situation in which the NNSD corresponds to a
Berry-Robnik parameter $q=0.83.$ The failure of spacing distributions
interpolating between those describing chaotic and regular systems in the
limit of near integrability is common in different RMT descriptions based on
generalized statistical mechanics, e.g. in \cite%
{evans,toscano,nobre,abul,bertuola,abul1}. As mentioned above, the
superstatistical random-matrix ensemble is base invariant. The Hamiltonian
of an integrable system, on the other hand, by definition, has a well
defined complete set of eigenstates, which constitutes a preferred basis.
The RMT approach to mixed systems cannot depart far from the state of chaos
without breaking base invariance.

\begin{figure}[ptb]
\begin{center}
\includegraphics[width=\columnwidth,height=5cm]{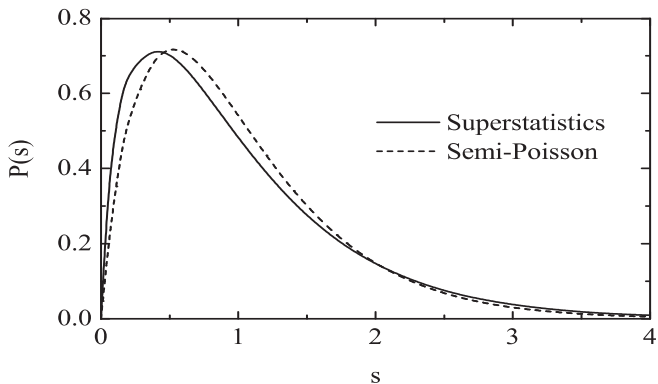}
\end{center}
\caption{ The superstatistical NNSD $p_{\,\text{Inv}\,\protect\chi^{2}}(0,s)$
compared with the semi-Poisson distribution $P_{SP}$.}
\label{fig_6}
\end{figure}

The variance $\sigma^{2}$ of the NNSD is often regarded as a one-parameter
interpolation between chaos and order because it monotonically increases
from $(4/\pi-1)\cong0.273$ for the Wigner distribution to 1 for the
Poissonian. At $\nu=0$, the superstatistical distribution in Eq.~(\ref{P2})
becomes%
\begin{equation}
p_{\,\text{Inv}\,\chi^{2}}(0,s)=\frac{\pi^{2}}{4}sK_{0}\left( \frac{\pi}{2}%
s\right)
\end{equation}
and has a variance $\sigma^{2}=\left( 16/\pi^{2}-1\right) \cong0.621$. This
value is slightly larger than 0.5 which is exactly the variance of the
semi-Poisson distribution%
\begin{equation}
p_{\text{SP}}(s)=4s\exp(-2s).
\end{equation}
The semi-Poisson distribution was suggested to describe a narrow
intermediate region between insulating and conducting regimes exemplified by
the Anderson localization model \cite{evangelou}, with the two limiting
cases being described by Poisson and Wigner statistics, respectively. It was
introduced to mimic new seemingly universal properties in certain classes of
systems, in particular, being characteristics of the \textquotedblleft
critical quantum chaos\textquotedblright. Figure~\ref{fig_6} compares the
limiting 
superstatistical distribution, \ $p_{\,\text{Inv}\,\chi^{2}}(0,s)$, with the
semi-Poisson distribution. We see from the figure that the two distributions
are quite similar in shape; the superstatistical distribution seems to just
have passed the semi-Poissonian before reaching its final shape for $\nu=0$.
Therefore it is worth in the present context to use the semi-Poisson
statistics as a reference distribution marking the limit of validity of the
base-invariant random-matrix description of mixed systems.

\section{Resonance strength}

In the preceding section we have seen that superstatistics yields a
theoretical description of statistical properties of the eigenvalues of a
quantum system with mixed classical dynamics, which is of similar quality as
existing ones, like e.g. the Brody and the BR ansatz. The great advantage of
the method of superstatistics is, that we can similarly apply it to
calculate the distribution of the eigenvector components of a Hamiltonian
describing a mixed system in which the spectrum is composed of subspectra
associated with levels following Poissonian and Wigner statistics \cite%
{sust3}. For a chaotic system the squared eigenvector components follow the
Porter-Thomas distribution 
\begin{equation}
P_{\text{PT}}(t)=\sqrt{\frac{{\eta}}{\pi y}}e^{-{\eta}t},  \label{PTT}
\end{equation}
where ${\eta}$ is a constant parameter related to the mean value \cite%
{barbosa}. In \cite{sust3} superstatistics of the transition matrix elements
was introduced by representing the transition-intensity distribution as a
superposition of Porter-Thomas distributions with different values for the
parameter $\eta$. Similarly, in a superstatistical description of the
squared eigenvector components of the Hamilton operator or equivalently the
partial widths of the resonances of a microwave resonator, the parameter ${%
\eta}$ in Eq.~(\ref{PTT}) is no longer considered to be a constant but
allowed to fluctuate according to a distribution $f({\eta})$. The
superstatistical distribution of the squared eigenvector components is then
given by%
\begin{equation}
P_{\text{Sst}}(t)=\int_{0}^{\infty}f({\eta})\sqrt {\frac{{\eta}}{\pi t}}e^{-{%
\eta}t}d{\eta}.  \label{PTTs}
\end{equation}
The parameter distribution ${f}({\eta})$ was determined in \cite{sust3}
using maximum-entropy arguments \cite{sattin}. There, the resulting
expression for the transition intensity distribution was fit to the
distributions of the experimental reduced transition probabilities in $^{26}$%
Al and $^{30}$P nuclei \cite{adamsx,shriner}. It fits the data much better
than a $\chi^2$~distribution with $\nu$ degrees of freedom, which is given
in Eq.~(\ref{P1}) with $\eta_0=1$ and has been proposed by Alhassid and
Novoselsky \cite{alhassid}\ and frequently used for describing the deviation
of partial width distributions from the Porter-Thomas distribution. The
superstatistical distributions obtained from Eq.~(\ref{PTTs}) for a $\chi^2$%
, inverse $\chi^2$ and log-normal distribution $f(\eta)$ (see Eqs.~(\ref{P1}%
)-(\ref{P3})), respectively, read%
\begin{equation}
P_{\chi^2}(\nu,t)=\frac{\Gamma\left( \frac{\nu+1}{2}\right) }{\sqrt{\left(
\nu-2\right) \pi t}~\Gamma\left( \frac{\nu}{2}\right) }\left( \frac{t}{\nu-2}%
+1\right) ^{-(\nu+1)/2},  \label{PT1}
\end{equation}%
\begin{eqnarray}
P_{\,\text{Inv}\,\chi^{2}}(\nu,t)&=&\frac{2^{(1+\nu)/2}(\nu+2)}{\nu~\sqrt{\pi%
}\Gamma\left( \nu/2\right) }\left[ t(\nu+2)\right] ^{(\nu-1)/4} \\
&\cdot& K_{(1+\nu)/2}\left( \sqrt{t(\nu+2)}\right) ,  \notag  \label{PT2}
\end{eqnarray}
and 
\begin{equation}
P_{\text{\,logn}}(v,t)=\frac{1}{\sqrt{2t}\pi v}\int_{0}^{\infty}\frac {1}{%
\sqrt{\eta}}\exp\left[ -\eta t-\frac{\ln^{2}\left( 2\eta e^{-v^{2}/4}\right) 
}{2v^{2}}\right] d\eta.  \label{PT3}
\end{equation}
Unfortunately the latter integral could not be evaluated analytically.

\begin{figure}[ptb]
\begin{center}
\includegraphics[width=\columnwidth]{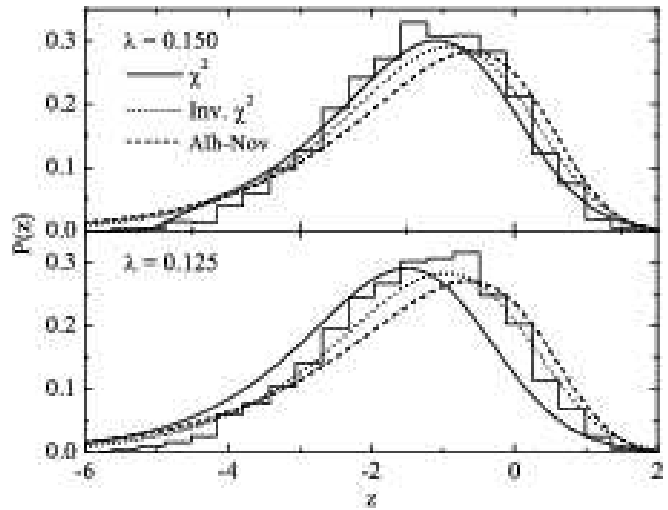}
\end{center}
\caption{ The resonance strength distributions in the mixed Lima\c{c}on
billiards compared with the predictions of different superstatistics. The
solid lines represent the $\protect\chi^{2}$, the dashed lines the inverse $%
\protect\chi^{2}$ distribution, while the short-dashed line corresponds to
the case in which the strengths have a $\protect\chi^2$~distribution.}
\label{fig_7}
\end{figure}

In \cite{dembo}, a new statistics, the resonance strength distribution was
introduced. For each resonance, a measurement of the transmission of
microwave power from one antenna to another provides the product of the two
partial widths related to the antenna \textquotedblleft
channels\textquotedblright. This product has been named the strength of the
resonance with respect to the transmission between the antenna channels. In
Ref. \cite{dembo} the strength distribution was investigated experimentally
for all three Lima\c{c}on billiards. Altogether four antennas were attached
to each of the microwave billiards and the transmission spectra between
pairs of antennas (a,b) were measured for all six possible antenna
combinations. An analytic expression for the strength distribution of a pair
of partial widths with distributions $P_{\text{a}}$ and $P_{\text{b}}$ is
obtained as%
\begin{equation}
P\left( y\right) =\int_{0}^{\infty}P_{\text{a}}(t_{\text{a}})P_{\text{b}}(t_{%
\text{b}})\delta(y-t_{\text{a}}t_{\text{b}})dt_{\text{a}}dt_{\text{b}}.
\label{strdis}
\end{equation}
The distribution of the partial widths of a chaotic microwave billiard is
well described by the Porter-Thomas distribution Eq.~(\ref{PTT}), and the
corresponding resonance strength distribution is given as 
\begin{equation}
P_{\text{GOE}}(y)=\frac{K_{0}\left( \sqrt{y}\right) }{\pi\sqrt{y}}.
\label{strdisGOE}
\end{equation}
More generally, if the partial width distribution is a $\chi^2$~distribution
as proposed by Alhassid and Novoselsky, then Eqs.~(\ref{strdis}) and (\ref%
{P1}) aith $\eta_0=1$ yield 
\begin{equation}
P_{\text{Alh-Nov}}(\nu,y)=\frac{2^{1-\nu}\left( \nu\sqrt{y}\right)
^{\nu}K_{0}\left( \nu\sqrt{y}\right) }{y~\Gamma^{2}\left( \nu/2\right) }.
\label{strdisg}
\end{equation}
The superstatistical resonance-strength distributions are obtained by
substituting the corresponding partial-width distributions (Eqs.~(\ref{PT1}%
)-(\ref{PT2})) into Eq.~(\ref{strdis}). If the parameter $\eta$ has a
log-normal distribution this leads to a double-integral which is not easy to
calculate. For the case of a $\chi^{2}$ distribution (see Eq.~(\ref{P1})),
the resonance-strength distribution is given by 
\begin{eqnarray}
P_{\chi^2}(\nu,y)&=&\frac{1}{\left( \nu-2\right) \pi\sqrt{y}~\Gamma
^{2}\left( \nu/2\right) }\left( \frac{y}{(\nu-2)^{2}}\right) ^{-(\nu +1)/2}
\\
&\cdot & G_{22}^{22}\left( \frac{y}{(\nu-2)^{2}}\left\vert 
\begin{array}{c}
1,1 \\ 
\frac{\nu+1}{2},\frac{\nu+1}{2}%
\end{array}
\right. \right)  \notag  \label{strdis1}
\end{eqnarray}
where $G_{pq}^{mn}\left( x\left\vert 
\begin{array}{c}
a_{1},...,a_{p} \\ 
b_{1},...,b_{q}%
\end{array}
\right. \right) $ is Meijer's $G$-function \cite{grad,luke,wolfram}. On the
other hand, if the parameter ${\eta}$ has an inverse $\chi^{2}$
distribution, one obtains 
\begin{eqnarray}
P_{\,\text{Inv}\,\chi^{2}}(\nu,y)&=&\frac{(\nu+2)^{2}}{4\pi\nu^{2}~%
\Gamma^{2}\left( \nu/2\right) } \\
&\cdot &G_{04}^{40}\left( \frac{1}{16}(\nu+2)^{2}y\left\vert -\frac {1}{2},-%
\frac{1}{2},\frac{\nu}{2},\frac{\nu}{2}\right. \right).  \notag
\label{strdis2}
\end{eqnarray}
Figure~\ref{fig_7} compares the experimental resonance strength
distributions for the 
two mixed Lima\c{c}on billiards, expressed as functions of $z$ = log$_{10}$ $%
y$, with the corresponding distribution given in Eqs.~(\ref{strdisg}), (\ref%
{strdis1}) and (\ref{strdis2}). The best-fit values of the parameters are
given in Table\ref{table2}. Both cases demonstrate the superiority of the
superstatistical inverse $\chi^{2}$ distribution.

\section{SUMMARY}

Superstatistics has been applied to the study of a wide range of phenomena
from turbulence to topics in econophysic. RMT of the Gaussian ensembles is
among these. In its application, the variance of the distribution of the
matrix elements is chosen as a parameter whose distribution is obtained by
assuming a suitable form or applying the principle of maximum entropy. In
this paper we use the time-series method to show that the spectra of mixed
systems have two correlation scales as required for the validity of the
superstatistical approach. The time-series analysis also shows that the best
choice of the superstatistical parameter distribution for a mixed system is
an inverse $\chi^{2}$ distribution. We computed the corresponding NNSD of
the energy levels and compared it with the spectrum of two microwave
resonators of mushroom-shaped boundaries and two of the family of Lima\c{c}%
on billiards, which exhibit mixed regular-chaotic dynamics. The agreement is
found to be similar to that with all the other considered well-established
distributions including the celebrated BR distribution. The method of
superstatics also provides a description of statistical properties of the
eigenfunctions of a system with mixed classical dynamics. Thus, the
resonance-strength distributions for the Lima\c{c}on billiards could be
analyzed. The agrreement with the experimental resonance-strength
distributions is better than with that derived from the Alhassid-Novoselsky $%
\chi^2$~distribution of transition intensities.

\begin{acknowledgements}
We are grateful to T. Guhr for stimulating discussions on the
subject of this work. One of us (A.Y. A-M) is grateful to the
quantum chaos group for the hospitality at the Institute of
Nuclear Physics of the TU Darmstadt, where part of this work was
done. This work has been supported by the DFG within the SFB 634 
and by the Centre of Research Excellence in Nuclear and Radiation Physics
at the TU Darmstadt.
\end{acknowledgements}

\end{document}